\documentclass[seceq,supplement]{ptptex}
\usepackage{graphicx}
\usepackage{amsmath}
\usepackage{bm}
\usepackage{url}

\newcommand\eref[1]{(\ref{#1})}
\newcommand\figref[1]{Fig.~\ref{#1}}
\newcommand\tabref[1]{Table~\ref{#1}}

\newcommand\vek[1]{\bm{#1}}  

\title{
  Visualizing a large-scale structure of production network\\
  by N-body simulation%
}

\author{
  Yoshi \textsc{Fujiwara}%
}

\inst{
  NiCT/ATR CIS Applied Network Science Lab
}

\abst{
  Our recent study of a nation-wide production network uncovered a
  community structure, namely how firms are connected by
  supplier-customer links into tightly-knit groups with high density
  in intra-groups and with lower connectivity in inter-groups. Here we
  propose a method to visualize the community structure by a graph
  layout based on a physical analogy. The layout can be calculated in
  a practical computation-time and is possible to be accelerated by a
  special-purpose device of GRAPE (gravity pipeline) developed for
  astrophysical N-body simulation. We show that the method
  successfully identifies the communities in a hierarchical way by
  applying it to the manufacturing sector comprising tenth million
  nodes and a half million edges. In addition, we discuss several
  limitations of this method, and propose a possible way to avoid
  all those problems.%
}

\begin{document}

\maketitle

\section{Introduction}\label{sec:intro}

Production network, or supplier-customer network, in economics refers
to a line of economic activities in which firms buy intermediate
goods from ``upstream'' firms, put added-value on them, and sell the
goods to ``downstream'' firms.

We recently studied a nation-wide production network comprising a
million of firms and millions of supplier-customer links in Japan by
applying recent statistical methods developed in complex networks (see
Ref.~\citen{fujiwara2008lss}). In particular, we found that firms cluster
into tightly-knit groups with high density in intra-groups and with
lower connectivity in inter-groups, and that this community structure
has sectoral and regional modules.

In order to verify the intra-group and inter-group connectivities, we
used a method of visualization of the entire manufacturing sector by a
graph layout based on a physical simulation. Such a visualization of
the large-scale network would be not only useful to check the community
structure, but also for visualizing several influences that are taking
place on the network including chain of bankruptcy, propagation of
demand, and influence of the variation in commodity-price.

In this paper, we fully explain the method of visualization, show that
the resulting layout successfully identifies the communities. In
addition, we discuss about the limitations of our method and also a
possible solution. \S\ref{sec:data} briefly describes the definitions
of nodes and links as firms and supplier-customer relationships, and
the network to visualize. \S\ref{sec:commun} compactly shows the
method of community extraction and its results.  Then, in
\S\ref{sec:force}, we explain our formulation of N-body simulation for
graph drawing so as to show how the graph layout is related to the
detected communities in a hierarchical way. The graph layout is an
energy-based placement of nodes. This method alone has several
limitations, as discussed in \S\ref{sec:disc}, where we shall also
propose a strategy of how to avoid them. \S\ref{sec:sum} summarizes
the paper.

\section{A nation-wide production network}\label{sec:data}

Let us say that a directional link is present as $A\rightarrow B$ in a
production network, where firm $A$ is a supplier to another firm $B$,
or equivalently, $B$ is a customer of $A$. While it is difficult to
record every transaction of supply and purchase among firms, it is
also pointless to have a record that a firm buys a pencil from
another. Necessary for our study are data of links such that the
relation $A\rightarrow B$ is crucial for the activity of one or both
$A$ and $B$. If at least one of the firms at either end of a link
rates the other firm as most important suppliers or customers, then
the link should be listed.

Our dataset for supplier-customer links is based on this idea. Tokyo
Shoko Research, Inc., one of the leading credit research agencies in
Japan, regularly gathers credit information on most of active firms.
In the credit information of individual firm, suppliers and customers
that are most crucial for each firm are listed up to the maximum of 24
firms respectively. We assume that the links playing important roles
in the production network are recorded at either end of each link as we
describe above, while we should understand that it is possible to drop
relatively unimportant links from the data.

We have a snapshot of production networks compiled in September 2006.
In the data, the number of firms is roughly a million, and the number
of directional links is more than four million. The set of nodes in
the network covers essentially most of the domestic firms that are
active in the sense that their credit information is required. See
Ref.~\citen{fujiwara2008lss} for the study of statistical properties
in the large-scale structure of the production network including
scale-free degree distribution, disassortativity, correlation of
degree to firm-size, low transitivity and so on, also for the relation
to chains of bankruptcies taking place on the network.

The global connectivity shows that basically all industries are highly
entangled among each other within the weakly or strongly connected
component. Yet the connectivity alone does not tell any modular
structure, namely how dense or sparse the stream of production is
distributed depending on industrial or geographical groups, which we
shall focus in the next section.

\section{Community extraction}\label{sec:commun}

Nodes in complex networks often cluster into tightly knit groups with
high density of intra-group edges and a lower density of inter-group
edges. Such a heterogeneous structure is called {\it community
  structure\/} of the network. The production network obviously has
dense and sparse streams of production distributed inhomogeneously
depending on industrial or geographical groups.

We focus, in this paper, on the manufacturing sector with 0.14 million
firms, in order to visualize the sector's modular structure by
excluding other dominant sectors including wholesale and retail trade,
which obviously have a different role in the stream of production from
the core of manufacturing sector. We regard the entire manufacturing
sector as an undirected graph, whose largest connected component turns
out to be consisted of 138,103 nodes (firms) and 421,893 edges
(supplier-customer links).

We use the method of maximizing modularity, introduced by
Ref.~\citen{newman2004fad} and implemented for large-scale graphs in
Ref.~\citen{clauset2004fcs} as a greedy optimization. While
considerable studies have been conducted to develop various methods for
community extraction, we use the modularity optimization for its clear
interpretation in terms of statistical hypothesis. Let $e_{pq}$ be the
fraction of edges in the network that connect nodes in group $p$ to
those in group $q$, and let $a_p\equiv\sum_q e_{pq}$, $b_q\equiv\sum_p
e_{pq}$. Then modularity $Q$ is defined by
\begin{equation}
  Q=\sum_p(e_{pp}-a_p b_p)
  \label{eq:def_Q}
\end{equation}
which is the fraction of edges that fall within groups, minus the
expected value of the fraction under the hypothesis that edges fall
randomly irrespectively of the community structure. The method is
formulated as an optimization problem to find a partition of nodes
into mutually disjoint groups such that the corresponding value of $Q$
is maximum.

We apply the method of community extraction to the undirected subgraph
whose nodes consist of only firms in the manufacturing sector.
The resulting modularity \eref{eq:def_Q} exceeds 0.55, which is
considered to indicate strong community structure. The number of
extracted communities exceeds a thousand, whose sizes range from a few
to more than 10,000. From the data on the attributes of the firms, we
found that many of those small communities are each located
in same geographical areas forming specialized production flows. An
example is a small group of flour-maker, noodle-foods producers,
bakeries, and packing/labeling companies in a rural area.

Because of a potential problem in the method of community extraction
by modularity optimization (see
Refs.~\citen{fortunato2007rlc,kumpula2007lrc}), we checked the
structure of detected communities by constraining modularity
optimization on each single community, especially for those with
relatively large community-size. Indeed, five large communities
exceed 10,000 in each size, being possibly subject to this problem of
resolution. After checking the sub-communities in the stated way, we
obtained the communities as tabulated in \tabref{tab:community} (see
Ref.~\citen{fujiwara2008lss} for more details).

{\small
\begin{table}[tb]
  \caption{\label{tab:community}%
    Communities extracted for the subgraph composed of manufacturing
    firms as nodes (about 0.14 million). Modularity optimization was
    recursively done for largest communities to obtain the
    sub-communities, ten of which are shown here. In each of them are shown ten
    firms with largest degrees are listed with names, major groups
    (primary/secondary/tertiary, if any in this order) of industrial
    sectors (see Ref.~\citen{jsic}), and sub-community sizes.}
\begin{tabular}{p{2cm}p{11cm}}
  annotation &
  firms (major groups; primary/secondary/tertiary), $\ldots$\hfill[community-size] \\
  \hline
  heavy indry. &
  Mitsubishi Heavy Industries (30/26),
  Kawasaki Heavy Industries (26/30),
  Kobe Steel (23/25),
  Ishikawajima-harima Heavy Industries (30/26),
  Sumitomo Heavy Industries (26),
  Nippon Steel (23),
  Kubota Industries (30/27/23),
  Mitsui Engineering and Shipbuilding (30),
  Hitachi Zosen Shipbuilding (26),
  Sumitomo Metal Industries (23),
  $\ldots$\hfill[7,447] \\
  automobile &
  Honda (30/27),
  Nissan (30),
  Toyota Motor (30),
  Aisin (25/30/27),
  Mitsubishi Motors (30),
  Denso (30/27),
  Fuji Heavy Industries (30),
  Toyota Industries (30/26),
  Suzuki Motor (30),
  Isuzu Motors (30),
  $\ldots$\hfill[5,769] \\
  materials &
  Sumitomo Osaka Cement (22),
  Air-Water Industrial Gas (17/18),
  Kyowa Concrete (22),
  Hokukon Concrete (22),
  Marukin Steel Materials (23),
  Mitsubishi Construction Materials (25/22),
  Hinode Steel/Manhole (23/22),
  Nihon Kogyo Industrial (22/13),
  Lafarge Aso Cement (22),
  Maeta Concrete (22),
  $\ldots$\hfill[2,644] \\
  electronics(a) &
  Hitachi (28/29/27),
  Fujitsu (32/28),
  NEC (28/29),
  TDK (27/29),
  Oki Electric (28/29),
  Hitachi High-Technologies (31/26),
  Rohm Semi-conductors (29),
  Murata Electronics (27),
  IBM Japan (28),
  Japan Radio Communication Equipment (28/27),
  $\ldots$\hfill[3,082] \\
  electronics(b) &
  Matsushita (Panasonic) (27/31),
  Sharp (29/27/28),
  Sanyo (27/25),
  Panasonic Shikoku Electronics (29/27/28),
  Pioneer (27/28),
  Matsushita Battery (27),
  Sanyo Tottori (28),
  Matsushita Refrigeration (27/26),
  Kenwood (28),
  CMK Electronic Devices (29),
  $\ldots$\hfill[2,921] \\
  electronics(c) &
  Canon (28/26/31),
  Seiko Epson (28/29),
  Omron (27),
  Nikon (31/26),
  Ricoh (26/28),
  Fujinon Optics (31),
  Hoya Optics (31),
  Casio (26/31/28),
  Pentax Optics (31/28),
  Sony EMCS Electronic (27/28),
  $\ldots$\hfill[2,692] \\
  electronics(d) &
  Toshiba (27/28/29),
  Stanley Electric (27/26),
  Toshiba Lighting and Technology (27/26/29),
  Ushio Electric (25/27/26),
  Hamamatsu Photonics (29/27),
  Nippon Electric Glass (22),
  Toshiba Tec (26/27),
  GS Yuasa Industry (27/29),
  Iwasaki Electric (27),
  Topcon Electric (31),
  $\ldots$\hfill[2,320]\\
  \hline
\end{tabular}
\end{table}
}

Each firm is classified into one or more industrial sectors, and the
major-group classifications (2 digits; see Ref.~\citen{jsic})
Obviously a community contains those firms in closely related
industrial sectors.  The annotations --- heavy industries, materials,
automobile, etc. --- are made by such observation. This recursive
procedure reveals a hierarchical structure in the communities, as we
have actually done so for the communities of so-annotated
``electronics'' (a)--(d), which constitute a single community in the
first stage of optimization.

We remark that these large firms in a same community do not form a set
of nodes that are mutually linked in nearly all possible ways, or a
quasi-clique. Rather, with their suppliers and customers, they form a
quasi-clique in a corresponding bipartite graph as follows. A
supplier-customer link $u\rightarrow v$ for a set of nodes $V$
($u,v\in V$) can be considered as an edge in a bipartite graph that
has exactly two copies of $V$ as $V_1$ and $V_2$ ($u\in V_1$ and $v\in
V_2$). Those large and competing firms quite often share a set of
suppliers to some extent, depending on the industrial sectors,
geographical locations and so on.

For the case of electronics (a)--(d), those quasi-cliques are further
separated into groups. Namely, the suppliers belong to different
groups of industrial organization for historical development and the
so-called {\it keiretsu\/}, and/or are located in divided geographical
sectors. The sub-communities (a)--(d) can be considered as such
separate groups with mutually sparse links. The electronics (b), for
instance, are originated and developed in an urban area in the western
Japan, not in the eastern urban area of Tokyo, being different from
the group (a).

In order to check the intra-group and inter-group connectivities, we
resort to visualization of the entire manufacturing sector by a graph
layout based on a physical simulation as shown in the next section.

\section{Energy-based placement}\label{sec:force}

We assume that the network is represented as an undirected graph
neglecting the direction of links, where nodes are firms and edges are
supplier/customer links. In general, desirable criteria for a readable
graph layout are:
\begin{enumerate}
\item adjacent nodes are placed closely,
\item nodes spread well in the layout, and
\item layout should be static as an equilibrium state.
\end{enumerate}
These criteria are very intuitive ones. Indeed, intuitive enough to
give us a physical analogy in {\it Nature\/}. A spring for an edge
does not allow adjacent nodes to stay too far apart nor too close, so
the analogy of the spring force will help one to satisfy the criterion
1.  Similarly, if the nodes are given Coulomb charges, say plus, then
they will spread moderately in a space satisfying the criterion 2. If
one combines such spring and repulsive forces appropriately, then the
system can have an equilibrium or quasi-equilibrium state, which might
be used as the solution for the criterion 3.

Indeed, the seminal paper\cite{eades1984hgd} is based on such a
physical modeling in graph drawing. There exists a huge literature of
graph drawing, in which physical analogies have been successfully
employed (see Ref.~\citen{dibattista1998gda,kaufmann2001dgm} for
review).

Our method is based on the following physical modeling. Each node $i$
is represented as a point-particle of mass $m_i$, and has a Coulomb
charge $q_i$. Repulsive forces are exerting between all pairs of nodes
by assuming that $q_i$ has a same sign for all $i$.  Each edge for
adjacent nodes $i$ and $j$, denoted by $\langle{i,j}\rangle$, is
replaced with a spring whose natural length is $\ell_{ij}$ and a
spring constant is $K_{ij}$. The spring force obeys the Hooke's law,
it attracts (or repels) the adjacent nodes if it is extended (or
compressed) than the natural length. In addition, a frictional force
is exerting on each node being proportional to its velocity with
coefficient $\gamma_i$. The friction decreases the total energy of the
system, while nodes are not staying still, so one has a certain
quasi-equilibrium state having locally minimum-energy.

The position $\vek{x}_i$ of each node $i$ obeys the equation of motion:
\begin{equation}
  m_i\frac{d^2\vek{x}_i}{dt^2}=
  \text{Coulomb}+\text{Spring}+\text{Frictional}\ ,
  \label{eq:eom}
\end{equation}
where
\begin{align}
  & \text{Coulomb} = C\,q_i\sum_{j\not=i}^N q_j\,
  \frac{\vek{x}_i-\vek{x}_j}{|\vek{x}_i-\vek{x}_j|^{3}}\ ,
  \label{eq:f_coulomb}\\
  & \text{Spring} = \sum_{\langle{i,j}\rangle}^M K_{ij}
  \left(|\vek{x}_i-\vek{x}_j|-\ell_{ij}\right)\ ,
  \label{eq:f_spring}\\
  & \text{Frictional} = -\gamma_i\frac{d\vek{x}_i}{dt}\ ,
  \label{eq:f_friction}
\end{align}
and $N$ is the number of nodes (particles), $M$ is the number of edges
(springs), and $C$ is the constant of Coulomb interaction. $\vek{x}_i$
is assumed to be a position in two or three-dimensional Euclidean
space. In the two-dimensional case, the expressions of Coulomb and
spring forces may not be validated as ``physically'' correct, e.g.
Coulomb force would be inversely proportional to the distance, but not
its square, but we do not care about them here.

For a sparse graph, $M\ll N^2$ by definition, so the computational
cost is largest for the calculation of Coulomb interaction being of
the order $O(N^2)$. A straightforward calculation does not work for
large $N$, and several techniques have been invented in different
disciplines of computational physics. If one replaces the charge with a
mass, and the Coulomb interaction with a gravitational interaction,
then one has an N-body calculation in gravitation. A well-known
technique is the Barnes-Hut algorithm,\cite{barnes1986hnl} \ which
reduces the number of particles in the summation \eref{eq:f_coulomb}
by replacing many forces from distant particles with a force from the
center-of-mass of them dividing the space recursively into quadtrees
(for two-dimension) or into octtrees (for three-dimension). The
computational cost becomes $O(N\log N)$. Another well-known technique
is a fast multipole method.\cite{greengard1987fap} \ We shall use the
Barnes-Hut algorithm\footnote{%
  Although the fast multipole method is of $O(N)$ in the calculation
  cost being preferred to the tree method theoretically, it may not be
  practically better for our purpose here, because it scales as
  $O(N\,p^2)$ for the order $p$ of multipole expansion whereas the
  tree code scales as $O(N\log Np)$ (see Ref.~\citen{makino1998fmm}).
  Elongated and distorted configuration of nodes in the graph layout
  would require larger $p$ than the case in which Coulomb screening
  (plus and minus charges) is present. We remark that a detailed study
  should be done for comparison, however.%
}.

The calculation can be further accelerated by a special-purpose device
of hardware, GRAPE (gravity pipeline), frequently used in
astrophysical N-body simulations (see Ref.~\citen{makino1998sss} for
review). Although the device was invented for the force summation for
gravity, it can be used directly for the summation \eref{eq:f_coulomb}
of Coulomb force as well. The other calculation for the forces of
spring and friction, integrating the equations of motion, and so on
are performed on a general-purpose computer. We used a GRAPE-7 (model
600, K\&F Computing Research Co.), which has the 120 pipelines on a
board operating at 100MHz clock cycle (power consumption 30W), so 12G
floating-point operations per sec. Assuming that a pair of interaction
needs 30 floating-point operations, the peak speed amounts to be
360Gflops. Such a system has a superior performance in computational
efficiency {\it per\/} cost of money, as demonstrated by a
finalist of the Gordon Bell Prize in Ref.~\citen{kawai2006ans}.

\begin{figure}
  \centering
  \includegraphics[height=0.80\textheight,bb=76.2002 225.0 549.0 774.0]{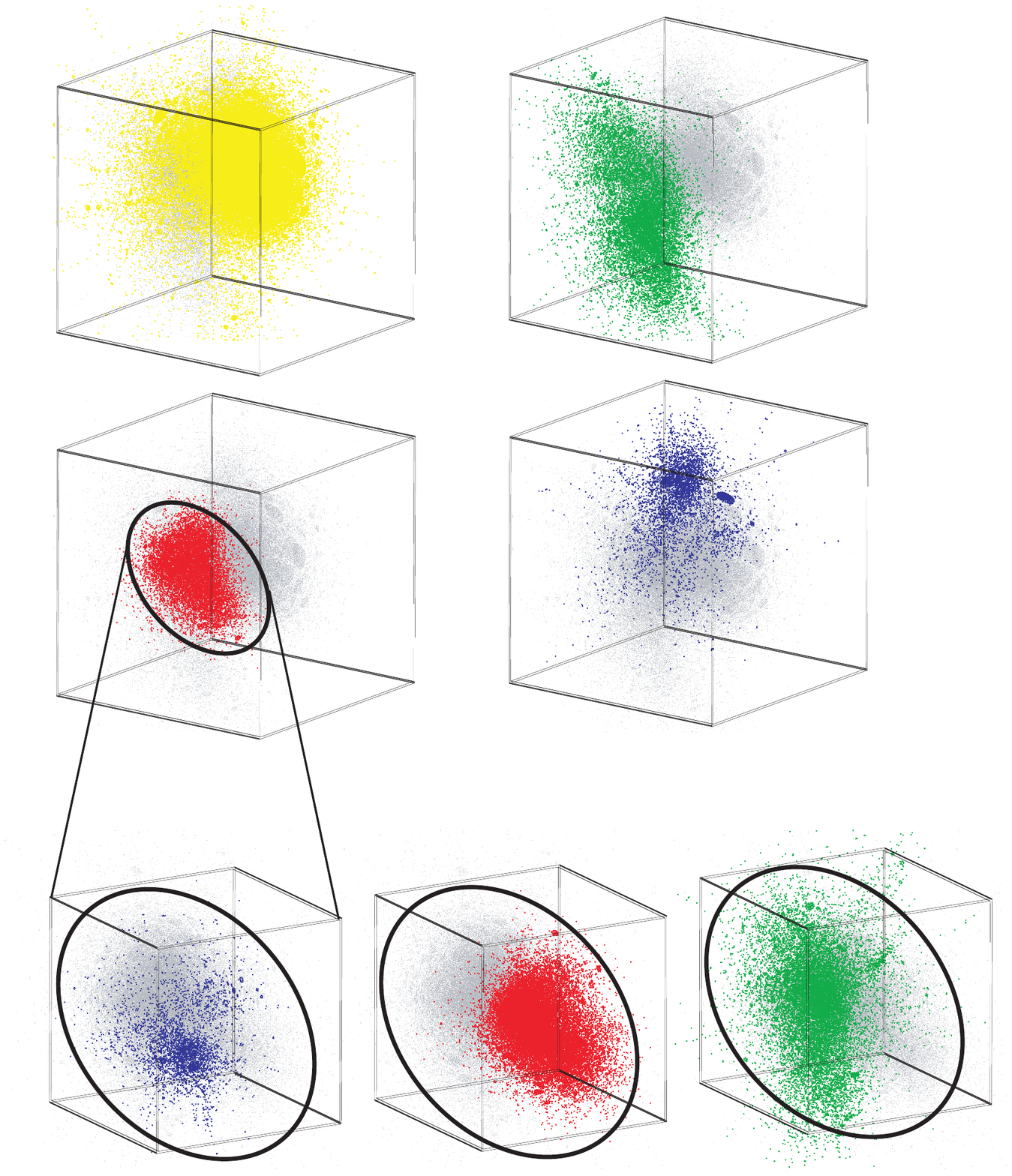}
  \caption{%
    (Color online) Three-dimensional visualization of the entire
    manufacturing sector with 138,103 nodes for firms and 421,893
    edges for supplier-customer links.  The nodes are depicted as
    black or gray dots, and the edges are omitted for visibility.
    Different communities are shown in each drawing by using black
    dots for nodes belonging to a same community. From top to middle
    and from left to right, the communities are annotated respectively
    as heavy industries, materials, electronics and automobile. In the
    bottom are shown the sub-communities of electronics (a)--(c).  See
    \tabref{tab:community}.%
  }
  \label{fig:lgl}
\end{figure}

We apply our method to the largest connected component in the subgraph
for the manufacturing sector described in \S\ref{sec:commun}. The
parameters are $m_i=1.0$, $q_i=2.7\times10^{-3}$, $\gamma_i=2.7$,
$K=8.4\times10^{-2}$, $\ell=7.2\times10^{-6}$, where the spring
constant and its natural length are assumed to be constant. The
computation takes only 10 minutes or so on a general-purpose computer
with CPU Intel Core2 Duo, 2.4GHz, and 2GB of physical memory with a
board of the device GRAPE-7 (see above for the spec) on a PCI-X slot.

Our result is depicted by \figref{fig:lgl} where the position
$\vek{x}_i$ of each node is represented as a point in a
three-dimensional space. In each drawing, a community is represented
by black nodes that belong to the community, whereas the other nodes
are gray dots. The largest communities are shown from top to middle in
the figure, which are further analyzed by the community extraction
based on modularity optimization as described in
\S\ref{sec:commun}. This recursive detection of communities give
sub-communities, three of which are shown in the bottom of
\figref{fig:lgl}, corresponding to the electronics (a), (b) and (c)
annotated in \tabref{tab:community}.

By the very nature of spring modeling, the nodes tend to cluster dense
subgraphs, most notably for communities. Because the drawings in
\figref{fig:lgl} do not explicitly use the information of community
structure, the graph layout by our proposed method alone successfully
represent the clustering of nodes in its visualization. Moreover, if
one zoom into a subgraph, one can visually identify sub-communities in
a hierarchical way as we have shown here.

\section{Discussion}\label{sec:disc}

Our proposed method directly solves the equations of motion to seek
for a quasi-equilibrium state of nodes placement. As described in the
beginning of \S\ref{sec:force}, there are various methods based on
physical analogies in the literature of graph drawing. Those methods
are roughly classified into two categories, namely force-directed
placement and energy-based placement. Our method belongs to the
latter. See chapter~4 in Ref.~\citen{kaufmann2001dgm}.

A frequently used as a routine by force-directed placement is due to
Fruchterman-Reingold.\cite{fruchterman1991gdf} \ A force-directed
placement means that each update of locations for nodes is simply
proportional to the total force. This implies that one solves an
equation of motion as if the mass were negligible in a certain limit,
or in other words, as if there were no inertia. A recent usage of
GRAPE in a similar context as ours in Ref.~\citen{matsubayashi2007fdg}
is based on a force-directed placement.

As an energy-based placement, Kamada-Kawai\cite{kamada1989adg} is
frequently used in many implementations. This algorithm does not
employ any repulsive forces between nodes, but uses springs of
different length and strengths between {\it every\/} pairs of
nodes. It assumes that the spring's natural length is given by the
shortest-path distance between nodes, while the spring constant is
inversely proportional to the squared distance. A locally minimum
energy is obtained by using a modified Newton-Raphson method. A recent
usage of the Barnes-Hut algorithm can be found in Ref.~\citen{quigley2000pfg}.

Our methods differ at least in the aspects that the direct solution
for the equation of motion, not force-directed, is performed and that
repulsive and spring forces with dissipation are used as described in
the previous section. Although one may want to compare those different
methods in a unified way for a given set of test datasets, it is
actually the case that different methods can be ideal for different data
because networks have various properties in the topological features
relevant to graph layout. So it would not be very practical to attempt
a systematic comparison. Rather, in our opinion, a good criterion in
choosing an algorithm would be then feasibility and
simplicity of a method even for large-scale graph, and we consider
that our proposed method is quite scalable (to astronomically large
scale) and simple enough to implement the code and the device (as done
in astrophysical context).

Thus the N-body simulation with the Barnes-Hut algorithm accelerated
by GRAPE is satisfactory in generating the overall visualization and
identifying community structure. Nevertheless, there are several
obvious drawbacks in the present method.

\begin{description}
\item[Initially random configuration]:\hspace{1em}%
  Nodes are placed at random positions in the initial configuration,
  and then it starts the calculation of dynamics for them. An amount
  of calculation is necessary for the dynamical system to settle in a
  quasi-stable state. If one has a good guess of what is the overall
  configuration, most of the computation could be skipped in the
  calculation of dynamics.
\item[Entangled final configuration]:\hspace{1em}%
  As a result, the quasi-stable state often yields a twisted and
  entangled configuration of edges between nodes, even if the network
  has a global structure such as a backbone and a tree, it is
  difficult to find such a structure as a quasi-stable state.
\item[Non-uniqueness]:\hspace{1em}%
  Starting from a different random configuration, one ends up with
  another configuration as a final layout. This is highly undesirable,
  because different layouts are generated from a same network in
  different simulation of the dynamics.
\item[Clustering]:\hspace{1em}%
  The present method involves spring-forces. Such a spring-type method
  geometrically cluster dense subgraphs by its very nature. Although
  we have shown that our method actually does so in the final layout
  to a certain degree, it would be desirable that a more explicit
  mechanism is included in the dynamics so that nodes in a same
  cluster are located nearby.
\end{description}

\begin{figure}[htbp]
  \centering
  \includegraphics[width=0.98\textwidth,bb=185 218 656 377]{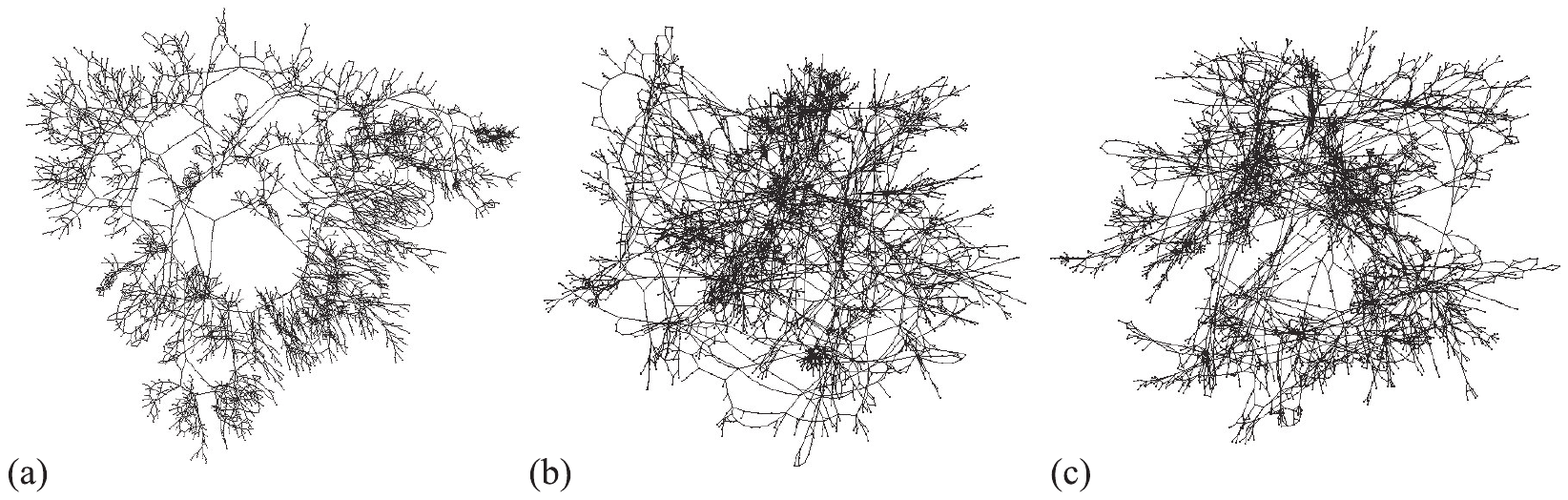}
  \caption{%
    (a)~US power-grid with 4,941 nodes and 6,594 edges. (b) and (c)
    are quasi-stable configurations resulting from two random
    configuration as an initial state of the same network.}
  \label{fig:powergrid}
\end{figure}

The first three points can be illustrated in \figref{fig:powergrid}.
The power-grid shown in \figref{fig:powergrid}~(a) has a backbone
which can be observed as a big circular structure when one ignores
many small branches and trees hanging off from it. However, starting
from two random initial configurations, one ends up with (b) and (c) as
typical results, which are highly entangled configurations, and also
look quite different from one another. If one uses energy-based layout
throughout from the starting configuration to the final, it might be
difficult to avoid these problems.

However, these problems could be solved by an additional idea which
follows from the goal of graph-drawing itself. Although it is beyond
the scope of the present paper to elaborate it, let us mention about a
basic idea. The elaboration will be published elsewhere.

Suppose a similarity measure $\delta_{ij}$ is defined for any pair of
nodes $i$ and $j$. We assume that it satisfies that
$\delta_{ij}\geq0$, $\delta_{ij}=\delta_{ji}$, and $\delta_{ii}=0$.
Any graph naturally has such a measure, namely, shortest-path lengths
between nodes. The goal of graph-drawing is, in general, to configure
each node $i$ at a location $\bm{x}_i$ in a conventional Euclidean
space typically of two or three dimensions such that the distances
between any pairs of nodes approximate the similarities as well as
possible:
\begin{equation}
  |\bm{x}_i-\bm{x}_j|\approx\delta_{ij}\ .
  \label{eq:mds_idea}
\end{equation}
As a result, similar nodes are located nearby, while dissimilar nodes
will be put away from each other. The most simple case would be that
the similarity is a distance between cities, then the resulting
configuration of cities must be equal to exactly a geographical
location of the cities on a map. The problem of graph-drawing is much
more difficult, but can be stated as how to solve the equation
\eref{eq:mds_idea} under a certain criterion for the approximation so
as to realize the low-dimensional embedding and representation.

It turns out that this problem has exactly the same essence in {\it
  multi-dimensional scaling\/} (MDS) (see Ref.~\citen{borg2005mms} for
example). Indeed, the energy-based placement by Kamada and
Kawai,\cite{kamada1989adg} \ one of the most frequently used spring-layouts,
can be interpreted as a kind of MDS (see Chapter~4 in
Ref.~\citen{kaufmann2001dgm} and also references therein). This
interpretation can lead one to the following idea. If one could solve
the equation \eref{eq:mds_idea} by using one of the MDS methods that
is applicable to a large number of nodes, then one can use the
solution as an initial configuration for the dynamics which then
follows exactly in the same way shown in this paper\footnote{%
  This is actually the method by which \figref{fig:powergrid}~(a) was
  obtained.}%
. Moreover, by appropriately designing the similarity measure to
include clustering of nodes (the fourth problem above), then it would
be possible to resolve those problems.

Therefore, we conclude this discussion to say that our approach could
be augmented by a better initial configuration, which might be given
potentially by an MDS solution to perform a practically fast
visualization in the same way as in this paper overcoming the problems
discussed here.

\section{Summary}\label{sec:sum}

In order to visualize communities in a nation-wide production network,
studied in the paper,\cite{fujiwara2008lss} \ we propose an N-body
simulation of Coulomb repulsive forces between nodes, Hook's spring
forces along edges and dissipation for relaxation. We showed that the
method successfully identifies the communities in a hierarchical way.
The calculation for the graph layout is done in a practical
computation-time and is possible to be accelerated by a
special-purpose device of GRAPE (gravity pipeline) for astrophysical
N-body simulation.  In addition, we discuss the limitations of the
method, and argue that all the problems could be solved by using an
appropriate calculation of initial configuration of nodes in a
multi-dimensional scaling.

\section*{Acknowledgements}

We would like to thank Yuji Fujita (NiCT) for collaboration and his
interactive system of GRAPE and {\tt Ruby}, and also Atsushi Kawai
(K\&F Computing, Inc.) for technical advice and support on GRAPE-7
model 600 and his library.

%



\begin{thebibliography}{99}

  

\bibitem{fujiwara2008lss}
Y.~Fujiwara and H.~Aoyama,
\newblock {\em {Large-scale structure of a nation-wide production network}},
\newblock available from {\tt http://arxiv.org/abs/0806.4280}.

\bibitem{newman2004fad}
M.~E.~J. Newman,
\PRE{69,2004,066133}.

\bibitem{clauset2004fcs}
A.~Clauset, M.~E.~J. Newman, and C.~Moore,
\PRE{70,2004,066111}.

\bibitem{fortunato2007rlc}
S.~Fortunato and M.~Barthelemy,
\JL{Proc.~Natl.~Acad.~Sci.~USA,2007,36}.

\bibitem{kumpula2007lrc}
J.~M.~Kumpula, J.~Saram{\"a}ki, K.~Kaski, and J.~Kert{\'e}sz,
Eur.~Phys.~J.~B~\andvol{56,2007,41}.

\bibitem{jsic}
Japan Standard~Industrial Classification,
\newblock \url{http://www.stat.go.jp/index/seido/sangyo/index.htm},
\newblock Rev.~11, March 2002.

\bibitem{eades1984hgd}
P.~Eades,
\JL{Congressus Numerantium,42,1984,149}.

\bibitem{dibattista1998gda}
G.~Di~Battista, P.~Eades, R.~Tamassia, and I.~G.~Tollis,
\textit{Graph Drawing: Algorithms for the Visualization of Graphs}
(Prentice Hall, Upper Saddle River, 1998).

\bibitem{kaufmann2001dgm}
M.~Kaufmann and D.~Wagner (Eds.),
\textit{Drawing Graphs: Methods and Models},
Lecture Notes in Computer Science no.~2025,
(Springer-Verlag, Berlin, 2001).

\bibitem{barnes1986hnl}
J.~Barnes and P.~Hut,
\JL{Nature,324,1986,446}.

\bibitem{greengard1987fap}
L.~Greengard and V.~Rokhlin,
\JL{J.~Comput.~Phys.,73,1987,325}.

\bibitem{makino1998fmm}
J.~Makino,
Bulletin of the Japan Society for Industrial and
Applied Mathematics~\andvol{8,1998,277}.

\bibitem{makino1998sss}
J.~Makino and M.~Taiji,
\textit{Scientific Simulations with Special-Purpose Computers}
(John Wiley and Sons, Chichester, 1998).

\bibitem{kawai2006ans}
A.~Kawai and T.~Fukushige,
\textit{\$158/GFLOPS astrophysical N-body simulation with reconfigurable
  add-in card and hierarchical tree algorithm},
in Proceedings of the SC06 (High Performance Computing, Networking,
Storage and Analysis) CDROM, 2006.

\bibitem{fruchterman1991gdf}
T.~M.~J.~Fruchterman and E.~M.~Reingold,
\JL{Software\ --\ Practice and Experience,21,1991,1129}.

\bibitem{kamada1989adg}
T.~Kamada and S.~Kawai,
\JL{Information Processing Letters,31,1989,7}.

\bibitem{matsubayashi2007fdg}
T.~Matsubayashi and T.~Yamada,
International Journal of Electronics, Circuits and Systems~\andvol{1,2007,116}.

\bibitem{quigley2000pfg}
E.~Quigley and P.~Eades,
\textit{FADE: Graph drawing, clustering, and visual abstraction},
in Proceedings of Graph Drawing 2000, Lecture Notes in Computer
Science no.~1984, pp. 197--210.

\bibitem{borg2005mms}
I.~Borg and P.J.F. Groenen,
\textit{Modern Multidimensional Scaling: Theory and Applications}
(Springer-Verlag, Berlin, 2005).

\end{thebibliography}
\end{document}